# Towards Requirements for a Demand Side Response Energy Management System for Households

## Research report, June 2019

In association with the EU-funded REPLICATE (Bristol) Smart Homes initiative and in collaboration with HoSEM and RES research projects

**By Caroline Bird and Ruzanna Chitchyan**

Department of Computer Science, University of Bristol, Bristol, BS8 1UB

**Abstract:** Demand response is considered to be one of the key means through which peak energy demand could be ameliorated. This report presents a requirements elicitation exercise (undertaken in collaboration with the Bristol City Council, UK) to elicit the requirements that a smart appliance automation service for domestic energy demand response management must address to be accepted by the households. The study comprises of an interview study with 28 householders, the findings from which were validated through two co-design workshops.

## Contents

# 1. Introduction

**The purpose of this report is to inform the design of household energy management service trials in the context of the Bristol REPLICATE (REnaissance in PLaces with Innovative Citizenship And Technology) Smart Homes project.**

The University of Bristol (through the collaboration of the EPSRC Household Supplier Energy Market (HoSEM) and Refactoring Energy Systems (RES) projects) has been working with the Bristol City Council REPLICATE team and participants in the Smart Homes project in the Easton, Ashley and Lawrence Hill wards of Bristol to explore the idea of smart energy management systems linked to Smart appliances supplied by the REPLICATE project.

The aim of this research is to understand home energy and appliance use and likely responses to an automated energy management system. We wanted to understand people's concerns, the constraints within which a system might work for them, drivers for engagement and the appropriate provision of information.

# 2. Context

In order to address climate change issues, energy use must be more efficient and based on generation from renewable sources. In the UK, progress is being made and electricity is increasingly coming from clean renewable sources such as wind and solar - these are different from dirty fossil fuel-based power as their supply is intermittent and requires a different approach to management and use. A further issue are the peaks in energy use which occur during early mornings (between 6am – 10 am) and evenings (4pm-8pm). Because of these, the grid needs to have sufficient generation facilities to accommodate periods of high demand. These peaks may also not match peak generation from renewables. **Demand-side response (DSR) services** respond to the issue of peaks of energy use to help people move their use of energy from these peak times to other periods in the day when energy is more plentiful. By managing peak demand, the grid can avoid unsustainable new generation investments and reduce the use of fossil fuel back-up generation, which cost money as well as causing increased environmental harm.

## 2.1 Background

The UK Department for Business, Energy and Industrial Strategy (BEIS) commissioned a report in 2017 on the potential of DSR in the context of small energy users [1] and we draw on the 'rapid evidence assessment' (REA) [2] element of this work for the purpose of our report. In particular, we are interested in their research question: '*What are the key factors affecting consumer engagement in terms of: recruitment, level of response and persistence*?'

Key findings are listed in the bullet points below:

- **Financial and environmental benefits** are main motivations for enrolment in DSR programmes but environment is generally insufficient on its own
- **Trust, risk and complexity are key factors affecting consumer engagement**
  **Trust** comes in 2 forms:
    o firstly in relation to who is setting up the system and why
    o Secondly during use of the system if users experience problems or there is a lack of transparency for example in relation to pricing or automation
    o Both of these can be addressed by the provision of appropriate, transparent and timely information, possibly through trusted and independent intermediaries

- **Complexity** relates to ease of use and clarity:
    - Automation can reduce effort unless setting up the system is difficult, once set up it needs to be both easy to use and easy to over-ride
    - Pricing structures need to be clear and low **risk** – ie users gain savings by switching to low price timings rather than incurring higher costs by not switching
- **Household routines**, or lack of them, affect the response to DSR – households with more flexibility and more time in the home are more likely to respond, and appliances which have less routines in usage are more flexible and responsive to demand shifting. Demand shifting that fits with or does not affect existing routines works best.

Demand response trials have recently been carried to explore community engagement in energy management, for example:

- Greenwich 'energy heroes' [3] used app alerts and rewards to encourage participants to shift energy use to times of lower demand
- Scottish and Southern's 'Solent Achieving Value from Efficiency' (SAVE) [4] programme used smart technologies in four different interventions trialling engagement and demand shifting

With the advent of Smart Meters, energy suppliers are able to develop variable pricing, 'Agile Octopus' [5] from Octopus Energy uses half-hourly prices, released daily and based on the wholesale price of electricity.

## 2.2 REPLICATE (REnaissance in PLaces with Innovative Citizenship And TEchnology)

REPLICATE is a European research and development project that aims to deploy integrated energy, mobility and ICT solutions in city districts. The city of Bristol, as part of a consortium with San Sebastián and Florence was awarded €25 million in October 2015 to create integrated smart city solutions to tackle urban problems such as traffic congestion, poor air quality and unsustainable energy use. The project has multiple elements both in Bristol and in its partner cities. In Bristol, the project focuses on the inner-east area of the city to deploy energy efficiency, sustainable mobility and connected smart digital services.

This research relates solely to the 'Smart Homes' element of domestic energy management. During 2018, 150 households in Bristol were recruited and supplied with smart appliances (Washing Machines, Dishwashers and Dryers) connected to apps and data collection devices. The aim of the Smart Homes project is to explore how energy demand might be better managed and shifted to assist in smoothing grid demand, avoiding times of peak energy use and potentially to take advantage of renewable generation or advantageous off-peak tariffs, using automated energy management systems.

The key question to be addressed by this research is:

> *What requirements should an automated energy demand service support in order to be positively received and adopted by households at large?*

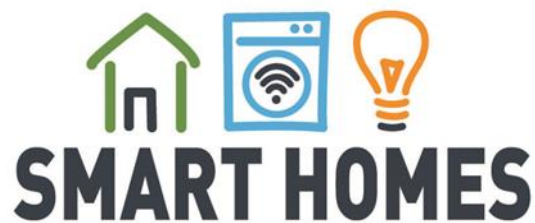

To address this question, we conducted a qualitative study to explore consumer responses within and beyond the REPLICATE project, addressing contextual variability, motivations and control issues.

# 3. Methodology

Our qualitative study sought to inform the design of the energy management service, using interviews and workshops to explore household limitations, expectations and acceptability of such services, from the perspective of the participating Bristol households. We also do not discuss the pre-interview trial, which is already reported in [6].

## 3.1 Interviews:

Semi-structured face-to-face interviews were conducted with a total of 28 interviewees during November 2018-Feb 2019. The interview questions were split into 3 sections:

i) initial knowledge and participant background details
ii) current practices of appliance and energy use
iii) reactions to the idea of automation and energy management

Interviewees were drawn from
a) households who had received smart goods from REPLICATE (total 16, **referenced as PR1-16)**
b) colleagues and friends external to the REPLICATE project (total 12**, referenced as PE1-12**).

In both sets of interviewees, whilst the ideal was to secure a representative sample, the reality was driven by who responded to appeals for participation. In both cases, diversity was actively sought and late respondents who seemed similar to those already interviewed were advised that their input was not required for the current phase of the project.

## 3.2 Workshops

Two workshops were held in March and April 2019 with a total of 17 external participants alongside members of the project team. They were designed to develop the interview findings and help inform the software and trial design for the energy management system. The findings from the interview study were also validated through exercises and feedback received at the workshops.

Exercises included:

a. Reflecting on personal routines and preferences to inform selection of time periods for running / not running appliances
b. Reflecting on how an energy management system might deliver maximum gains
c. Thinking through the sign-up and use of a system
d. Discussing the use of rewards or savings for personal or community gain
e. Validating findings from the interview study (both via set exercises, and as an additional brief questionnaire completion exercise in workshop 2)

## 3.3 Analysis

A Grounded Theory approach has been taken to analyse the interview findings and explore the theoretical grounding for developing our ideas. This approach firmly anchors our analysis and emerging outcomes in the data obtained from our participants, who in turn represent the future users of the system. Grounded Theory requires careful selection of data sources (i.e., participants), thorough coding of the data and ongoing re-examination of emerging themes. Through this approach, we are able to explore the complex socio-technical interactions of domestic energy use and automated energy management systems. Additional data was collected (via workshops) to strengthen/refute the ideas we proposed on basis of the interview data analysis.

# 4. Thematic Findings

Here we present the identified/validated themes from the interviews and workshop data. The theory building and validation results will be detailed in a separate paper.

## 4.1 Domestic practices and routines

Our interviews showed how participants went about their daily lives and how and when they used their various appliances.

Unsurprisingly, those interviewees who were in a part-time employment, had no small children or were retired had more flexibility in their appliance use times – they also tended to prefer being present when appliances were running. Larger households, those in full time work and those with children have less flexibility in their current practices but may also benefit more from the support of an energy management system. Use of appliances is thus connected to the (non-)existence of specific daily fixed routines:

- the unemployed and retired tended to have little routine, were more likely to use appliances during the day and had more preference to being present whilst appliances were running;
- Part-time employed used free days and weekends to do the washing;
- Full-time employed were the most likely to run appliances in evening peak periods, overnight or at weekends.

Within these variabilities, there were a number of typical use patterns, and common constraints to flexibility in relation to automation of the three types of REPLICATE appliances.

Whilst **washing machines** were in frequent use by all households, there were a number of typical limitations to their use:

- wanting to be present whilst an appliance is running (worries about malfunction);
- wanting to be present at the end of the cycle in order to empty it promptly for drying (worries about 'smelly' washing if left in the machine; or other people needing machine / interfering with washing e.g., with house-sharers or when hosting students);
- responding to clothing needs, particularly in relation to children (mess from baby, muddy sports kit, school uniform) ;
- drying practices relating to season – i.e., to hang washing outside in the *morning* in summer or near a radiator in the *evening* in winter;
- availability of drying space in the household;
- to not run at night if it was noisy and might disturb sleep/neighbours.

**Dryer** use patterns fell into 3 categories:

- ❖ regular, after every wash;
- ❖ regular after every wash when the weather was unsuitable for drying outside (especially in the winter);
- ❖ Occasional, when there's a particularly large amount of washing, for particular items, if something is needed in a hurry or to finish off air-drying.

Use of the dryer is thus closely linked to the washing machine, with most participants expecting to be able to move washing from washer to dryer promptly and for it to run without delay so that washing does not sit wet waiting to be dried. Users were happier to leave dried washing in the dryer, particularly the type supplied by REPLICATE where it gives gentle tumbling to prevent creasing. These condenser dryers are cooler than traditional ones, alleviating safety fears relating to over-heating.

**Dishwasher** use similarly fell into different categories:

- ❖ Daily, essential part of routine;
- ❖ Regularly, but waiting until it's full;
- ❖ Occasional, when there are guests.

The regular users tended to have regular patterns of use, generally after the evening meal, or sometimes overnight or after breakfast, with the latter more of a preference by those at home during the day.

Where it takes more than a single day to fill up, users were more likely to be concerned about the smell of dirty dishes and also more likely to use an 'intensive' wash' to clean off dried on food.

A particular concern relating to dishwashers, cited by 2 interviewees who knew of this occurrence, and others who had heard it advised, was that of a machine catching fire.

We also discussed **other appliances** that were important to the user and their potential for flexibility. These fell into four main categories:

1. Cooking appliances – this includes electric hobs, ovens, microwave, kettle, toaster, food mixer / blender. For most of these there is some small flexibility – for example the evening meal might be shiftable by 30mins but not 60mins. But, for many, the need / desire for sustenance meant that these were fixed uses. Slow cookers and bread makers meanwhile are generally used out of peak periods.
2. Entertainment – including games consoles, televisions and sound systems. A number of households have at least one sound system on all evening whilst others cited the needs of children in relation to consoles and television. Again, there is small shiftability but use is perceived to be relatively fixed in relation to daily routines
3. Household / work management – PC, electric heating, electric shower and water heating, lighting, hoover, iron
4. Personal – hairdryer, electric razor and toothbrush; chargers for laptop, phone, tablet etc.

## 4.2 What would automation mean to the user?

We asked participants about the likely benefits and possible problems that they could envisage using an automated system (e.g., connected to a smart-phone app) to manage their appliances. The main themes emerging from the interviews were further explored in the workshops and are discussed below.

1. <u>Convenience</u> takes a number of forms:

Optimisation of energy and price by handing over control so that the system can make best price and energy decisions within constraints set by the user. The user has the reassurance that use of clean energy is optimised and energy is being used at the cheapest time possible and, by working with an app, has confidence in what decisions are being made on their behalf with the app keeping them informed.

If the system works well, habits can changed over the longer term.

A further convenience is not needing to be physically present whilst appliances are turning on and running so that times of use have maximum flexibility and the system can choose the best time. The automation app may also make suggestions to help optimise appliance use.

Using an automated system also frees up time for other activities and reduces time wasted waiting for a cycle to complete. Washing is then ready when the user needs it. BUT it needs to be straightforward to use and effective in its outcomes.

> *'if we were to come home and the washing was already washed we'd just open the door and then hang it up, that would be very convenient' (PE1)*
>
> *'You'd want it to be smart enough to kind of look after itself' (PE2)*
>
> *'it would be amazing, but I think that's the thing, you'd have to have it done for you automatically, wouldn't you? Because remembering to do it all might be another added dimension onto busy lives' (PE10)*

2. <u>Energy efficiency is the core ambition,</u> for users this means better use of resources and better understanding of their energy use. Some points are discussed below:

The facts that the management system can *make decisions based on availability of renewable energy* and that will help the grid to be used more efficiently, is perceived as positive outcomes and motivators for continued engagement with automation. By evening-out energy use across the day, concerns about 'surges' or 'blackouts' are reduced. System and user flexibility helps make best use of resources to meet everyone's requirements. A future benefit is in the storage of energy to use at other times, and ensuring that electric vehicles are charged when renewable energy is plentiful.

*Feedback is important.* The system should show energy / $CO_2$ savings and potential benefits, enabling users to think differently about their consumption. Greater knowledge should mean that energy is used more consciously, thus enabling people to make changes and improvements to their own practices.

> *"Information about those peak times and how much energy is being used nationally at those times, and my average daily and weekly usage. Information like that would be good, because I haven't got a clue" (PR3)*

Other information might include how much energy is wasted using standby or leaving lights on. The system could suggest the best way to use appliances, give an overview of how energy has been used and how these compare with (relevant, comparable) others.

> *"I think there should me more sort of advice about how you can be more efficient and that sort of thing. And maybe sort of tailored to you, rather than just random." (PE7)*

Notifications from the app on what energy sources will be available and when it is sunny / windy will help in planning ahead to maximise use of renewables, with the system optimising use and showing outcomes. Users are more likely to use the longer but more efficient 'eco' settings if they are planning their washing/appliance use ahead.

Respondents hoped that appliances, as well as being more efficient, would be long lasting and repairable.

3. Financial benefits

Most interview respondents expected to see some sort of financial benefit and variable pricing is one approach that we explored in the workshops. Users need more information on possible savings and expect the system to support financially beneficial decision making with minimal input from the user.

> *"If you could physically see, wow, look how much it costs then, if I'm doing it at peak times, to say, for example, look how much it costs if it's on at three in the morning. I reckon if you could physically see that, you wouldn't mind at all whenever it was used it was cheaper" (PE10)*

> *"I think that is good the automation thing. I think that would be good, if overall it helped cut costs then I'm for it." (PR13)*

4. Reducing risk of appliance failure / malfunction – which was a concern of the interviewees

Malfunction might be through the appliance overheating or other problems whilst the appliance is in use remotely through the management system. The workshop participants suggested that the automation system might be able to inform users of issues (e.g., those that could cause malfunction) as well as help them maximise usage efficiency. Ideas included:

- Notifications of when to deep clean appliance and filters; detection of power surge or burnout; flooding prevention – the system would be able to turn off the device and send a notification to the user (e.g., via phone message);
- The system could assess if the appliance is being used most efficiently and advise on needed changes;
- Advice on appliance operation and possible imminent failure, recording cycles and having diagnostic capability;
- Setting optimisation (e.g., detecting washing load's type, weight, colour in order to optimise settings).

5. Helping to bring about bigger "smart" changes

Beyond domestic automation of particular appliances, users were also interested in how this connects to a bigger picture in terms of new 'smart' approaches to household and city management and how it might help us collectively understand better energy use. As so often mentioned, information is key – for example on when best to use energy with real-time updates, monthly summaries and neighbourhood comparisons. One participant suggested that at its core, 'home is about warmth, security, food and social space' – a whole house system which reinforces this will better support the use of energy services.

Understanding smart at a domestic level can also facilitate 'smart city' buy-in and connect smaller actions to bigger 'household' and 'city' activity. Uber is an everyday example but there are many others, such as flexible workspaces, real-time bus arrivals, traffic monitoring, hourly bike and car hire etc.

6. <u>Alienation from 'normal', the 'robots' are taking over and 'big brother is watching'</u> are all concerns relating to an automated energy / home management systems.

Whilst these concerns were recognised, users also suggested that energy management systems could become a 'new normal' and that it is necessary to go through some discomfort or alienation in order to make the big changes that are necessary to change energy use and address climate change urgency.

> *"It's like everything's a bit weird until it's like your new normal, so my normal is to wash in the mornings and chuck it straight into the tumble dryer and that works for me 'cause that's all I've ever done but like if there's a new, you know, I only do washing at 6 o'clock or whatever time it is, then that's the new normal then that's fine" (PR13)*

> *"We won't find it a problem because we'd adapt and think, well we can't use things at certain time so you'd use it when you could." (PR7)*

However, to attract a wide user base, a new system must be user-friendly and a positive experience, providing benefits to the user, and being quick to feel like 'normal' by matching to existing or flexible household practices and making it easy to make changes. This also means that help is available for set-up – e.g., via video or phone line. It must also be accessible to all, including older people, less tech-able and those with disabilities.

From the interviews, respondents were also concerned about how a whole-house system might not understand what is needed in the home and that it might switch off the wrong things if it was set to optimise energy efficiency and smooth demand. Household dynamics are complex things and each one is different.

---

Positive benefits:
- o Makes the best decisions without needing input, something else to do the thinking so that optimisation is achieved without too much effort
- o Efficiency, frees up time for other things
- o Optimisation (energy, financial) – feel good that energy is optimised, renewables are maximised, energy is used when it's cheapest
- o Provision of information means that users understand energy better in general
- o Feedback and information helps keep participants motivated
- o More conscious energy use will influence other areas of daily life
- o User friendly system will quickly become a 'new normal'

or, conversely, negative impacts:
- o loss of control
- o worries about malfunction, lack of trust
- o complexities of a household not understood by machines

---

## 4.3 Using an energy management system

Through the interviews and workshops, we explored how users might use appliances within an energy management system. We first asked them for their initial reactions to such a system and then asked how they would prefer to use it and what sort of controls and approaches would work for them, further developing these responses through the workshops.

1. Setting time preferences: using a 24hr clock face, participants worked through how they might set usage time-bands for each appliance. Key issues in relation to setting time preferences for operation of appliances:

- Dryer and dishwasher uses are easier to shift – particularly as users are mostly happy to leave them unemptied for longer periods. Washing machines are more complicated as the washing needs to be taken out and dried more promptly;
- With knowledge of peak usage times and variable pricing (when available), it is quite easy to avoid peaks (without necessarily being automated);
- Settings in appliance control apps need to have 7-day variations (as people have different routines across the days of the week) and should also respond to winter / summer differences;
- Some users preferred to manage their own appliance use using the time-delay button;
- Pricing information is necessary to facilitate selection of cheap time periods;
- By aligning to current practices – e.g., to switch on at 2pm and hang out as soon as you get back from work or at 5am so that it's ready when you get up – the system is most likely to be used regularly

2. Service delivery: We used the notion of user journey (as shown in Fig. 1) to help users think through the steps of using an energy management system and what they might find or need at each step. The steps can be grouped into four main phases, with their relevant issues discussed below:

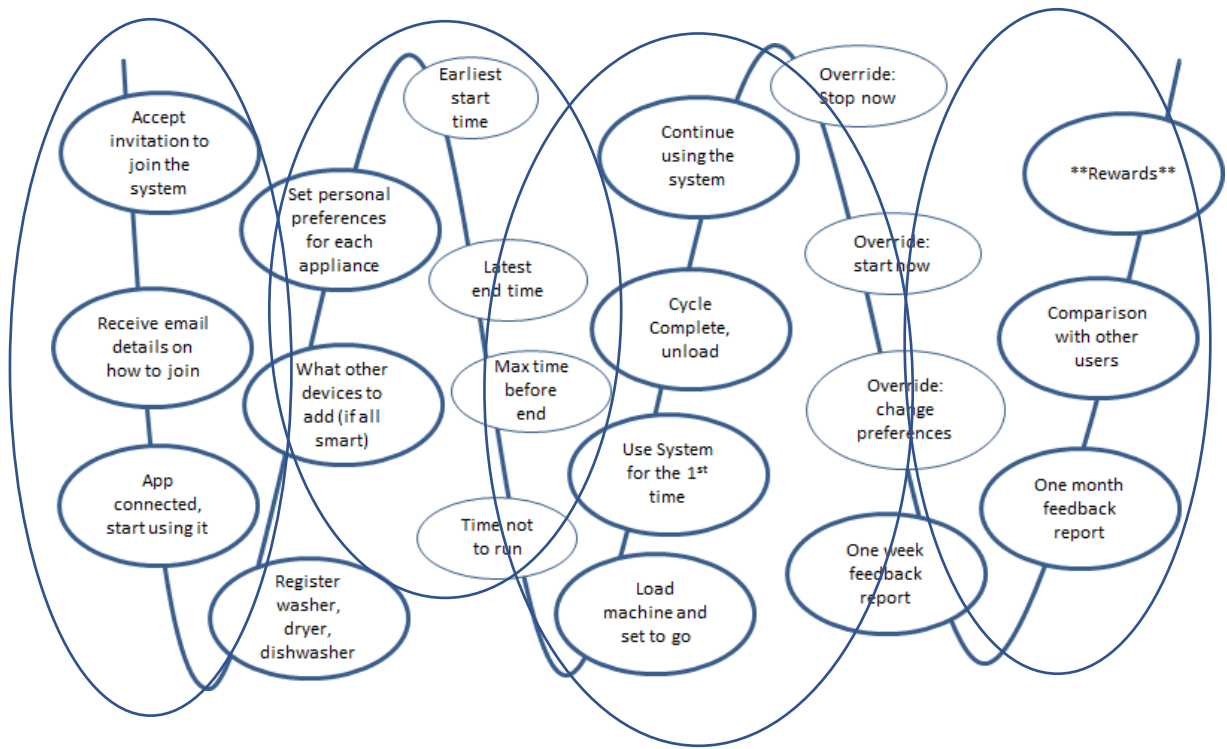

**Figure 1:** Demand-side response automation user journey

**(a) The Initiation phase**, when users find out about the new system, download the app and connect their devices to it:
- Clear advance information is needed and the app interface should look nice, be intuitive and work well immediately otherwise users will stop using it;
- Instructions for joining the system should be clear and 'non-techie' with lots of easily accessed support via a phone line, 'how-to' videos and physical demos;
- Users should know what's happening to their data.

**(b) Customising to own settings** should be easy to set up
- Users should understand how they (appliances and app) to easily setup preferences once and keep these standard, or how to change settings – preferences and overrides;

- Again, support should be available;
- How is the interface presented (e.g., sliders, checkboxes, etc.);
- How do users receive information on pricing and notifications on when it's best to run?
- Queries around multiple app users, house-sharers, and flexible / non-fixed routines.

**(c) Using the system** – and getting used to it
- Confidence grows with each successful use;
- Feedback is provided via the app so that users know what's happening;
- Single overrides should be easy without disrupting the overall regular settings;
- Need to fit with lifestyle to maintain usability.

**(d) Feedback and rewards**
- Weekly reports at the start are good to show what's happening and ensure that the system is working well. Users should then be able to set report frequency or to access information via the app / web;
- Monthly feedback and comparisons (with comparable users) will show use patterns over time and suggest improvements but users should be able to opt out of 'competition' mode if they don't want to make comparisons;
- Whilst financial benefits are good to see, rewards are not just financial; they include a sense of social / environmental responsibility so environmental impacts should also be part of the reports and feedback.

> Summary of key points for using an energy management system:
> ✓ Provision of clear and timely information
> ✓ Support and help – phone line, video, talk to others
> ✓ Easy to use, looks good, not too 'techie'
> ✓ It tells you what's going on in real time
> ✓ Feedback on how it is being used / could be used better
> ✓ Provision of comparisons with comparable others (or not – ability to switch on / off 'competition' mode)
> ✓ Recognition of financial, environmental and social outcomes

## 4.4 Motivations, rewards and pricing

Participants recognised different motivators for signing up to energy management and automation. Many of them said that **the environment** was a primary motivator – they want to be supportive of environmental protection as long as is practical within their household set-up. The households also would like to get feedback on the actual environmental impacts that they have, particularly if the DSM automation was a city-wide system.

Saving money was also a strong motivator although it was more likely to come second to the environment.

Some participants said that they were prepared to cope with some inconvenience:

> "I would sacrifice, you know, if something was maybe a bit disruptive or not perfect, then I would rather sacrifice that to reduce the energy consumption" (PE5)

> "I would be very willing to change the way that I do stuff, even at the cost of convenience or money, if that's really going to make a difference." (PE9)

In relation to **financial rewards**, we asked interviewees first, how important these were and then what type of pricing system might most incentivise them. Whilst participants were driven by environmental concerns, most said they would want financial savings too – even if they are quite small – in recognition of the initial disruption of getting to grips with a new system.

> "I think my main consideration would be if there was a financial incentive for use. I would consider a shift in use around my washing machine, dishwasher and the larger energy users where I could have some flexibility. I would consider using them differently." (PE11)

> "A financial incentive is really important" (PR4)

When asked how much would incentivise them, some respondents suggested a reduction of at least 10% on their current bills, others asked for at least £50 annual savings, while yet others stated that any saving would be gratefully received but it wasn't a main driver. Some suggested that any pricing or reward system would need to be designed so as not to penalise low-income families or those that had a relatively high power needs (e.g., having to use electric vs gas heating):

> "the disadvantaged area that we live in, penalising someone that's already disadvantaged that is just not great" (PR13)

Response to the idea of variable pricing was diverse. For instance, another respondent said:

> "Like you'd be penalised for using it between…, oh no. I wouldn't like that at all. If you've got children and they need things at certain times. It'd depend on.. [if] you're quite well off then that's fine, money's not a worry. But for poor families, that would. No that is terrible." (PR10)

Meanwhile others saw peak / off-peak pricing as a significant motivator to change their behaviours:

> "You have a peak price and a non-peak price. I think that would be an incentive definitely and get us thinking about energy quite differently I think and moving what you can away from the peak." (PE11)

Some respondents were quick to see the parallels with other sectors:

> "It's a bit inconvenient but I'll have the food delivered at 10 o'clock at night, oh I've got to stay up but it's £2 cheaper" (PE10)

> "That's kind of the standard model for a lot of things isn't it, like flights or you know, Uber" (PR2)

Suggestion for a rewards scheme type of approach was met with a mixed response, some respondents could see the benefits and thought it would work whilst others did not want to be made to feel guilty if they performed less well or failed to achieve a particular target:

> "You could build up points. I like the idea of building up points" (PR9)

> "…like your point cards at supermarkets. I think that would work with a lot of people." (PR12)

> "…then you get the guilt if you do it at the wrong time. And you're not going to get rewards because you've been naughty" (PE7)

In the workshop, participants strongly responded to the variable pricing details given during one of the exercises, choosing appliance running times that avoided the clearly marked 4-7pm peak times.

We also asked about whether **financial rewards** were a strictly personal thing or whether **sharing** into a community pot would be a good idea – especially as the savings are likely to be quite small. Interviewees generally responded positively to the idea of community benefit – although some did suggest that it very much depended on the community and how connected individuals felt.

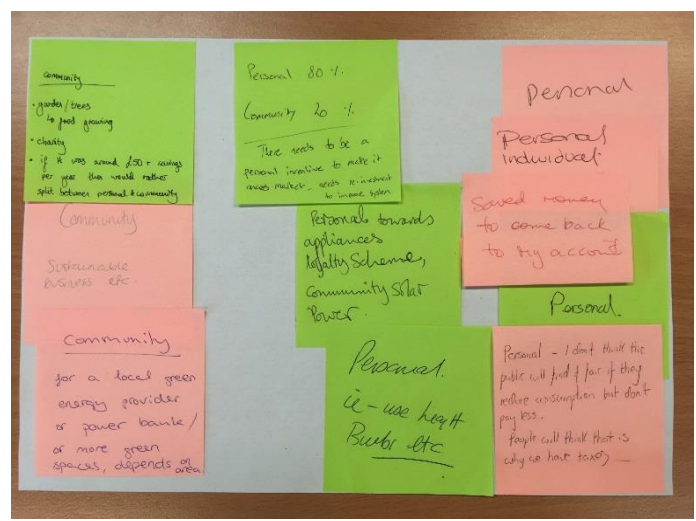

There was a split in the workshops between keeping and donating the savings. Overall, most people wanted to see some sort of financial gain from changing their routines and we saw a pretty even 3-way split between wanting all the savings, wanting at least half and donating it all.

Ideas for community projects included:
- Battery, local renewables, energy projects, green energy;
- Local parks, community centre, green spaces;
- Food growing – gardens / trees;
- Foodbanks, homeless shelters, addressing fuel poverty;
- School equipment, adventure playground, youth centres;
- Environment or sustainability projects and businesses;
- Set up a 'crowdfunder tariff'.

Finally, a number of participants recognised a degree of social responsibility to engage in managing energy better – both for supporting current users and future generations:

> "I don't mind looking stupid or old-fashioned but if you're looking socially irresponsible to not do it then that would motivate me." (PE9)

Motivations for participating in an <u>energy management system trial</u> vary, with some participants enthused and others more reluctant to cede control to the management system. These are some of the main drivers:
- Curiosity about how it might work, a desire to help develop new systems, interest in the research;
- Contributing towards creating a better future;
- Convenience, as it would help make good decisions for the users;
- Contributing towards addressing climate issues, environmental concerns;
- Playing one's part in society, social responsibility, contributing to community projects and wellbeing;
- Financial savings via a variable pricing tariff

**Overall, most participants were motivated by both environmental and financial outcomes .**

## 4.5 Access to and use of data

Data privacy was an underlying concern but it was also one that most participants didn't feel that they had much control over. A common response was to shrug and say, well it's all out there anyway, I've no idea who knows what about me or how to control it but there's nothing to hide anyway!

> "If I'm not doing anything wrong then I've got nothing to hide, right, that's my kind of mantra" (PR2)

When we delve more into ideas of data ownership and use, whilst some respondents had concerns over who was using data and for what. Many respondents accepted the idea that energy companies could hold and use detailed data – especially if it helped them to better manage supply and the grid – doing good for the planet:

> "for the common good, not for the private benefit" (PE1)

> "if the main aim is to save energy, it's about doing something good" (PE10)

However, to others the need to make data available might be a de-motivator for participation in energy management:

> "I can't say I'm enthusiastic about it, no. It would act as something of a demotivator. If I thought the whole thing was a really great idea, maybe I'd accept that but I don't really like the idea". (PE9)

Participants were less keen when it came to data sharing with private interests unless it helped develop new technologies, they did not necessarily want to be contacted or sold services as a result .

> "I guess, it's companies who are paying to get this data so that they can use it to improve their technology" (PR8)

> "I suppose as long as it doesn't then lead to sort of loads of calls saying "oh, we can sell this for you and sell you this and save you that", then yeah. Not being used for marketing" (PE6)

# 5. Discussion and conclusions

## 5.1 Summarised Requirements for DSR automation

Our earlier pilot study [6] suggested the following requirements for DSR automation service to households (in column 2)– this is updated below with data from this research and this table summarises some of our findings.

| Requirements Group | Initial Requirement | Updated overview |
|---|---|---|
| Control | R1. Manual Override | Manual override is viewed as essential to allow users to respond to the pressing needs of the household |
| | R2. Data Sharing | Most respondents in this later study were ambivalent about data sharing, regarding it as inevitable and largely outside their control already. Some expressed an interest in knowing who had the data and for what purpose |
| | R3. Traceability of Consumption | All participants expressed an interest in knowing / understanding more about their energy consumption and cost of different appliances. Visibility of automation was also expressed as a preference so that, at least initially, they could see that it was performing correctly, and trust would be developed. |
| Per-Device Automation | R4. Selective, per-device automation | Each appliance is used differently and has different constraints on use and the need for outcomes, hence per-device setting of use time and/or limits is confirmed as a preference. |
| | R5. Individual device context and profile | Additionally, users have different preferences and needs so personalisation is necessary. Examples: 1. some use dishwashers regularly at night whilst others worry about fire safety and want to be present and awake whilst it is in operation; 2. in family households with small children, one parent manages most of the washing whilst in shared households, those with lodgers or with adult children devices have multiple users. |
| Personalisation | R6. Set own goals | Drivers for engaging with automation services vary, although most suggest that the 'feel-good' knowledge of acting in a socially and environmentally responsible way is an underlying driver along with, ideally, some money saving. Opinions varied as to whether time-of-use pricing which would make clear the peak periods would be a better motivator than collecting 'rewards' for use shifting. |
| | R7. Define preferences per context, day, time | See R5 above |
| | R8. Maintain comfort | In the context of the later study, this refers to not being overly inconvenienced by the use of automation services. Most participants were accepting of some degree of change / disruption to current behaviours in recognition of outcomes for societal benefit. Some current behaviours are opportunistic and occur without much planning so a shift of some of these are unlikely to result in inconvenience beyond the need to undertake some modest forethought – e.g., to load a machine. |
| Default Participation | R9. Opt out vs opt in | In the context of REPLICATE, automatic opt-in might not be appropriate unless carful thought is given to how this might be operationalised – given the constraints and requirements listed above, automatic opt-in would need to be carefully managed and individual preferences used. |
| | R10. Social interconnection | In the current study, social interaction was not regarded as an important motivator except in the context of friends. In the case of REPLICATE, participants are drawn from across 3 wards areas of the city and are mostly not known to each other. Social learning cannot be discounted but, given the relative anonymity of participants currently, it might not be important |
| Education | R11. Inform on gains and losses | In the earlier pilot, participants were interested to know what they gain or lose through automation services. Now, participants did not expect to lose except in some minor inconveniences but were keen to be able to monitor what difference their shifting of energy demand might make. |
| | R12. Educate on contributions | Greater knowledge was a significant finding and participants generally are interested in what difference their actions make – in the context of energy, finances environment, and social responsibility. |

It is apparent from this work that, whilst many users are initially cautious in response to the idea of domestic energy automation, once the reasons are explained they become more accepting – particularly if they are supplied with good information on the potential impacts of making changes – these include environmental, financial and social benefits. Users should benefit from participation rather than losing by not being engaged – particularly those on low incomes.

Any future energy management system would need to be clear, easy to use and initially fit to current routines before facilitating the more substantial changes that would 'become the new normal' over time. Users expect to benefit from a new system – through increased convenience or financially to compensate for some initial inconvenience. Most accept some initial inconvenience or changing routines as an inevitable part of making the changes necessary to address climate and environmental issues.

The interviews allowed us to discuss home energy use and energy automation in some depth, with the workshops opening up these findings in order to help structure a future system. The co-design process that we have undertaken through the workshops is important to build consensus and really understand how potential users might respond to a new system.

## 5.2 REPLICATE - specific outcomes / recommendations

Whilst much of the discussion above is relevant to the setting up of any energy automation / management system, there are some specific recommendations in relation to the REPLICATE project.

Pricing / reward structure

We used the example of 'Agile Octopus' variable pricing in the workshop and this could be used for the basis of time variable pricing in the trial. For example: if an appliance uses 1.5kWh per usage and the time is shifted from a 25p/kWh time (4-7pm) to a 9p one (0-7am or 1-4pm) there is 16px1.5=24p saving. If the appliance is used 4 times per week the overall potential saving is 96p/week.

It's also 7.5kWh moved away from peak generation.

Possible pricing in bands reflecting Octopus' 'agile' tariffs over 3 months in 2019 might be:

|  |  | average | April | May | June |
|---|---|---|---|---|---|
| **8p bands:** | **Super off peak** | 00-07, 13-16h | 0-05, 13-16 | 00-05, 13-16 | 00-07, 13-16 |
| **10p bands:** | **Off peak** | 10-13, 21-24h | 05-07, 11-13, 22-24 | 5-7, 10-13, 2130-24 | 07-13, 21-24 |
| 13p bands | Shoulder | 7-10, 19-21h | 07-11, 19-22 | 7-10, 19-2130 | 19-21 |
| 24p band | peak | 16-19h | 16-19 | 16-19 | 16-19 |

| Time | 00 |  |  |  | 07 |  | 10 |  | 13 |  | 16 |  | 19 |  | 21 |  |
|---|---|---|---|---|---|---|---|---|---|---|---|---|---|---|---|---|
| tariff | 8p | 8p | 8p | 8/13 | 13 | 10 | 10/8 | 8 | 24 | 24/13 | 13/10 | 10 |

It might be helpful to make comparisons with the differential costs of trains, flights or delivery slots to demonstrate that this is how other markets work.

Participation

Drivers for initial participation:

- Free machine therefore some sense of obligation;
- Curiosity about how it might work;
- Want to help develop new systems;
- Being part of something new;

- Sense of achievement;
- Contributing towards addressing climate issues, playing your part.

What will help keep engagement?

- Information;
- Ease of use;
- Access to support and help;
- Doing it with a friend;
- Feedback on personal and system outcomes.

# 6. References and further reading

[1] *Realising the potential of demand-side response to 2025. A focus on Small Energy Users. Summary Report November 2017*. Department for Business Energy and Industrial Strategy, UK.

[2] *Realising the Potential of Demand-side response to 2025. A focus on small energy users. Summary Report; Rapid Evidence Assessment report. November 2017*. Department for Business Energy and Industrial Strategy, UK.

[3] Greenwich Energy Heroes: https://greenwichenergyhero.org/

[4] The Solent Achieving Value from Efficiency (SAVE) project: https://save-project.co.uk/

[5] Agile Octopus - A consumer-led shift to a low carbon future (2018) Octopus Energy https://octopus.energy/static/consumer/documents/agile-report.pdf

[6] *Eliciting Requirements for Demand Response Service Design to Households: A Pilot Study*. Ruzanna Chitchyan, Palvi Shah, Caroline Bird. EASE 2019, Copenhagen. URL: https://dl.acm.org/citation.cfm?id=3319355

# 7. Acknowledgements

Our thanks to the interviewees and workshop participants for their time and openness in responding to the questions and exercises, and to the REPLICATE team at Bristol City Council for their input to the design of the study and access to relevant project data.

**REPLICATE is funded by the European Union's Horizon 2020 programme**

**The following research is funded by the UK Engineering and Physical Sciences Research Council (EPSRC):**

a) EP/P031838/1 Household-Supplier Energy Market (HoSEM)
b) EP/R007373/1 Refactoring Energy Systems (RES)

# Appendix 1: FAQs derived from the interviews

Below is a list of the 'frequently asked questions' that can be drawn from the interviews, together with responses relating specifically to the REPLICATE project and appliances.

| | | |
|---|---|---|
| 1. | Who do I contact if I need help with the system? | See your 'useful contacts' sheet provided |
| 2. | What happens if my appliance or the EMS malfunctions? | If your appliance malfunctions please contact Samsung support. The energy management system can only work when smart control is switched on so simply switch this off if you have any concerns and contact the project team. |
| 3. | What happens if there is a power cut? | The energy management system will not function in this instance. |
| 4. | Do I need to do anything with my home wifi? | No. |
| 5. | My phone is quite old, will I still be able to take part? | Providing it is a smart phone and you have downloaded the smartthings app then yes. If you have a tablet then you can use the app on that also. |
| 6. | Can more than one person have the app? | Yes. |
| 7. | What do you mean by 'use times'? | The times at which your machine will be used. |
| 8. | Will the app tell me when my appliance will finish / has finished running? | If you set notifications for this then yes. |
| 9. | I want to be there when my appliance is running, will I be able to set it so that it does not run when I am out? | We recommend using the appliance when you are at home so please bear this in mind when setting your preferences. |
| 10. | I don't like my washing being left in the machine after it has run, will I be able to set it so that it finishes running when I am available to empty it? | Yes you will have total control when you set your start (and therefore end) time. |
| 11. | How long will I have to wait once I have loaded my appliance? | If you have loaded your appliance at a peak time, the energy management system may delay this so yes you would wait until the peak time ends. |
| 12. | Will you provide feedback on how the system is working for me / for everyone? | Yes you will be able to achieve a Gold, Silver or Bronze standard in terms of compliance with the energy management system preferences to work when energy is greener and cheaper to provide. |
| 13. | How will I be able to find out how much energy has been saved / shifted? | The amount of energy used will not differ. You will be able to see your level of compliance with the energy management system through a Gold, Silver, Bronze rating. |
| 14. | How do I know how much energy different settings use and how much they cost? | Please see your user manual for energy use guidance. The cost will depend on how much you pay for your electricity and will vary between household. |